\documentstyle[12pt,cite]{article}
 \def\bfe#1{{{\bf e}\left[#1\right]}}

\begin{document}

\addtolength{\baselineskip}{.3mm} \thispagestyle{empty}
\vspace{-1.5cm}
\begin{flushright}
INS-Rep-1128\\
December 1995
\end{flushright}
\vspace{15mm}
\begin{center}
  {\large\sc{$N=2$ heterotic string threshold correction,\\[2mm]
      $K3$ surface and generalized Kac-Moody superalgebra}}\\[18mm]
  {\sc Toshiya Kawai}\\[8mm]
  {\it Institute for Nuclear Study, University of Tokyo,\\[2mm]
    Midori-cho, Tanashi,   Tokyo 188, Japan} \\[27mm]
\end{center}
\vspace{1.5cm}
\begin{center}
  {\sc Abstract}
\end{center}
\vspace{5mm}
\noindent
We study a standard-embedding $N=2$ heterotic string compactification
on $K3\times T^2$ with a Wilson line turned on and perform a
world-sheet calculation of string threshold correction.  The result
can be expressed in terms of the quantities appearing in the two-loop
calculation of bosonic string. We also comment and speculate on the
relevance of our result to generalized Kac-Moody superalgebra and
$N=2$ heterotic-type IIA duality.


\newpage

\section{Introduction}

The work of Seiberg and Witten \cite{rSWi,rSWii}\ aroused much
enthusiasm in the investigation of $N=2$ supersymmetric Yang-Mills
field theories.  The pursuit of analogy \cite{rDKLL,rAFGNT,rKV} in
string theory has led to an exciting discovery of $N=2$ heterotic-type
IIA duality \cite{rKV}\ which is currently subject to intensive
scrutiny \cite{rFHSV,rKLT,rKLM,rVW, rAGNTi,rKKLMV,rAP,rAFIQ}.

To establish such duality it is important to know one-loop effects
in heterotic string compactified on $K3\times T^2$, especially its
moduli dependence.

One of the important quantities in probing such effects is string
threshold correction to the gauge coupling.  There is an extensive
literature on the subject
\cite{rMinahan,rKaplunovsky,rDKL,rOV,rFKLZ,rCFILQ,rAGN,rAGNT,
  rAT,rMSi,rBLST,rCLM,rMS,rKL}.  Roughly speaking, string threshold
correction is a logarithmic sum (with coefficients of gauge charges)
of masses of `heavy modes'.  Since heavy modes can be accidentally
massless at some points in the moduli space we encounter logarithmic
singularities at such points.  There is also constraint stemming from
the existence of an infinite number of light Kaluza-Klein particles in
the decompactification limit \cite{rCFILQ}.  With this information
along with the imposition of $T$-duality one can sometimes guess the
moduli dependence of string threshold correction.

On the other hand there is a beautiful world-sheet calculation
initiated by Dixon, Kaplunovsky and Louis \cite{rDKL}.  The connection
between the world-sheet calculation and the new supersymmetric index
\cite{newindex} was discovered in \cite{rAGN}.

In a recent interesting work, Harvey and Moore \cite{rHM}\ clarified
the role played by BPS states, extended the calculation of
\cite{rDKL}\ and pointed out the relevance of generalized Kac-Moody
(super)algebras \cite{rBorcherds,rNikulin,rGNi,rGNii}.

In this work we present the result of a world-sheet calculation
similar to those in \cite{rDKL,rHM}. We consider a standard embedding
heterotic string compactified on $K3\times T^2$ with a single Wilson
line turned on.  Employing the formula due to Gritsenko and Nikulin
\cite{rGNi}\ we show that the string threshold correction (for the
portion of our interest) can be written in terms of the expression
familiar in the two-loop calculation of bosonic string \cite{twoloop}.

Interestingly, the formula of ref.\cite{rGNi}\ we will use is blessed
with an interpretation of the denominator formula of a certain
generalized Kac-Moody superalgebra which is a special case of the more
general and geometrical constructions \cite{rGNii}\ associated with
the Picard lattices of algebraic $K3$ surfaces.  Since it has been
argued \cite{rAspinwall}\ that in string-string duality the Picard
group of $K3$ (the fiber of a $K3$ fibration) on the type IIA side is
related to the perturbatively visible gauge groups on the heterotic
side, it is tempting to consider that generalized Kac-Moody
superalgebras are the right framework in which to discuss
string-string duality, at least in the weak coupling regime of
heterotic string.

\section{$K3$ elliptic genus}

The calculation of string threshold correction involves elliptic genus
\cite{rSW, rWitteni,rAKMW}. In all likelihood, this has been known to
experts for a long time although may have not been stated explicitly.
In fact the appearance of elliptic genus is ubiquitous in the
literature \cite{rLerche,rDKL,rAGNTi,rAP}. A streamlined argument is given in
ref.\cite{rHM}.  The relevance of elliptic genus can be intuitively
understood as follows.  In the case of heterotic compactification on
$K3\times T^2$, string threshold correction can depend on $K3$ only
through coarse topological invariants like elliptic genus since the
$K3$ moduli fields belong to neutral hypermultiplets and hence do not
affect the physics of the vector multiplets.

Recall that (2,2) elliptic genus is defined by
\begin{equation}
  Z(\tau,z)=\mathop{Tr}\nolimits_{R,R}\,(-1)^{F}y^{(J_L)_0}
q^{L_0-\frac{c}{24}} {\bar q}^{{\bar L_0}-\frac{c}{24}}
\end{equation}
where $q=\bfe{\tau}$, $y=\bfe{z}$, $\bfe{x}=\exp[2\pi i x]$ and the
trace is taken over the (Ramond, Ramond) sector.  $J_L(z)$ is the
$U(1)$ current of the left $N=2$ algebra and $c=6$ for the $K3$ sigma
model.  For general properties of $N=2$ elliptic genera see
refs.\cite{rKYY,rKM}, the convention of which we will use in what
follows.

There are many expressions of the $K3$ elliptic genus corresponding to
many physical realizations \cite{rEOTY, rKYY, rKM}.  For our present
purpose, the following information suffices.

Due to the general properties of (2,2) elliptic genus \cite{rKYY}, it
can be shown \cite{rEZ} that the $K3$ elliptic genus $Z(\tau,z)$ has
an expansion
\begin{equation}
 Z(\tau,z)=\sum_{a \in{\bf Z}_{\ge 0},\, b \in {\bf Z}}\,c(4a-b^2)\,
 q^a y^b\,,
\end{equation}
where the coefficients $c(N)$ are  some integers.
As mentioned in \cite{rKYY}, $Z(\tau,z)$ can be organized as
\begin{equation}
  \label{egdecomposition}
  Z(\tau,z)=h_0(\tau)\, \theta_{0,1}(\tau,2z)+h_1(\tau)\,
\theta_{1,1}(\tau,2z)\,
\end{equation}
where
\begin{equation}
\label{theta}
   \theta_{0,1}(\tau,2z)=\sum_{b=even}q^{b^2/4}y^b, \quad
    \theta_{1,1}(\tau,2z)=\sum_{b=odd}q^{b^2/4}y^b,
\end{equation}
and\footnote{For explicit  expressions, see \cite{rKYY}.
Notice that the $h_0$ and $h_1$ there are different
 by a factor of $\eta(\tau)$.}
\begin{eqnarray}
  h_0(\tau)&=&\sum_{ N \equiv 0 \pmod 4} c(N)\, q^{N/4} \nonumber\\
&=&20 + 216\,q + 1616\,{q^2} + 8032\,{q^3} + 33048\,{q^4} +
117280\,{q^5}+\cdots\,, \\[4mm]
h_1(\tau)&=&\sum_{ N \equiv -1 \pmod 4} c(N)\, q^{N/4} \nonumber\\
&=&q^{-1/4}(2 - 128\,q - 1026\,{q^2} - 5504\,{q^3} - 23550\,{q^4} -
86400\,{q^5}+\cdots)\,.
\end{eqnarray}
Note in particular that
\begin{equation}
  c(0)=20,\quad c(-1)=2, \quad c(N)=0\ \mbox{if $N<-1$}\,
\end{equation}
thus
\begin{equation}
  Z(\tau,z)=2y+20+2y^{-1}+ O(q)\,.
\end{equation}
The Witten index is given by
\begin{equation}
  Z(\tau,0)=\chi(K3)=24\,.
\end{equation}

The decomposition (\ref{egdecomposition}) is related to the fact that
an $SU(2)$ affine Lie algebra (in the present case, at level one) is
contained in the $N=4$ superconformal algebra of the $K3$ sigma model
and $J_L(z)=2J_3(z)$ where $J_3(z)$ is the Cartan $U(1)$ current of
the affine $SU(2)$.  The relevant representation theory of $N=4$
superconformal algebra can be found in \cite{rET}.

\section{$N=2$ heterotic string threshold correction with a Wilson line}

We consider a heterotic string compactification on $K3\times T^2$ with
standard embedding.  The massless spectrum contains an $E_7$ vector
multiplet belonging to ${\bf 133}$ and 10 hypermultiplets belonging to
${\bf 56}$ of $E_7$.  (There are other 65 gauge neutral
hypermultiplets -- the 20 $K3$ moduli hypermultiplets and the 45 gauge
bundle moduli hypermultiplets --, but these hypermultiplets do not
come into our subsequent discussion.)

It is useful to review how this gauge symmetry arises from the
world-sheet point of view.  The gauge symmetry $E_7$ is entirely from
the left movers and is enhanced from $SO(12)\times SU(2)$ where the
$SO(12)$ symmetry is realized, in the fermionic formulation, by 12
free left gauge fermions $\lambda_a$ and the $SU(2)$ comes from the
internal symmetry of the left $N=4$ superconformal algebra of the $K3$
sigma model.  Thus for the vector multiplet, we have
\begin{equation}
  {\bf 133}\rightarrow ({\bf 32_c,2})+({\bf 1,3})+({\bf 66,1})
\end{equation}
and for the hypermultiplets,
\begin{equation}
  {\bf 56} \rightarrow ({\bf 32_s,1})+({\bf 12,2})\,.
\end{equation}
The free gauge fermions $\lambda_a$ and the $K3$ sigma model must be
in the same periodic boundary conditions, {\it i.e.\/} both in the
Neveu-Schwarz sectors or in the Ramond sectors.  We now focus on the
Ramond sector of the $K3$ sigma model.  Since the $\lambda_a$ in the
Ramond sector realize a level one $SO(12)$ affine Lie algebra of
spinor conjugacy classes, the relevant parts are $({\bf 32_c,2})$ for
the vector multiplet and $({\bf 32_s,1})$ for the hypermultiplets.
Therefore in the decomposition (\ref{egdecomposition}) of the $K3$
elliptic genus, the first term corresponds to the hypermultiplets and
the second term to the vector multiplet.  Note that $c(-1)$ and $c(0)$
correctly count, with a multiplicity two, the number of the vector
multiplet and hypermultiplets respectively.

In addition, there are three other abelian vector multiplets
associated with $S$, $T$ and $U$ where $S$ is the dilaton-axion, $T$
and $U$ are the K\"ahler and complex structure moduli of the torus
$T^2$. The graviphoton is contained in the graviton multiplet.

We will now consider a situation where a Wilson line is turned on so
as to break the above SU(2) gauge symmetry down to $U(1)$.  This
Wilson line modulus is denoted by $V$ which is a member of the vector
multiplet of the unbroken $U(1)$. Thus for $V\ne 0$ the vector
multiplet of $({\bf 32_c,2})$ becomes massive while the
hypermultiplets of $({\bf 32_s,1})$ are neutral with respect to the
$U(1)$ and remain massless.

We will investigate the dependence of the string threshold correction
on $T$, $U$ and $V$.  Since the Narain duality group $SO(2,3;{\bf Z})$
is isomorphic to $Sp(4,{\bf Z})$, it is  convenient to combine
these parameters into a $2\times 2$ matrix belonging to the Siegel
upper half plane of genus 2:
\begin{equation}
  \label{syuuki}
  \Omega=\left(
    \begin{array}{cc}
     T&V\\
     V&U
    \end{array} \right),\qquad  Im T>0,\quad Im U>0,
     \quad Y:=\det Im \Omega>0\,.
\end{equation}
Thus the threshold correction must be in some way expressed in terms of
automorphic forms of $Sp(4,{\bf Z})$.

We start with the integral
\begin{equation}
\label{thresh}
{\cal I}=
  \int_{{\cal F}}\frac{d^2 \tau}{Im \tau}\sum_{m_1,m_2\atop n_1,n_2}
\left( \sum_{b=even}q^{ \frac{1}{2}p_L^2}\bar q^{ \frac{1}{2}p_R^2}h_0(\tau)+
\sum_{b=odd}q^{ \frac{1}{2}p_L^2}\bar q^{ \frac{1}{2}p_R^2}h_1(\tau)
-c(0)\right)
\end{equation}
where $\cal F$ is the fundamental domain of the modular group of the
world-sheet torus and
\begin{eqnarray}
  \frac{1}{2}p_R^2&=&\frac{1}{4Y}\left\vert
m_1U+m_2+n_1T +n_2(TU-V^2)+bV
 \right\vert^2\\[2mm]
  \frac{1}{2}(p_L^2-p_R^2)&=&\frac{1}{4}b^2-m_1n_1+m_2n_2
\end{eqnarray}
with $m_1$, $m_2$, $n_1$, $n_2$ and $b$ running over integers.  The
subtraction in (\ref{thresh}) is to remove the logarithmic
singularities due to the massless hypermultiplets.

We closely follow the neat calculation done initially by Dixon,
Kaplunovsky and Louis \cite{rDKL}\ and recently extended by Harvey and
Moore \cite{rHM}.  As in these works, the calculation of ${\cal I}$
involves three contributions {\it i.e.\/} ${\cal I}={\cal I}_0+ {\cal
  I}_{nd}+ {\cal I}_{deg}$.

 Omitting the details the result is as
follows.

(1) zero orbit
\begin{eqnarray}
{\cal I}_0&=&\frac{Y}{ U_2}\int_{{\cal F}}\frac{d^2 \tau}{(Im \tau)^2}
         Z(\tau,0)\\
          &=&\frac{Y}{ U_2}\cdot\frac{\pi}{3}\cdot 24
\end{eqnarray}

(2) non-degenerate orbits
\begin{equation}
 {\cal I}_{nd}=-\log \prod_{k>0,l\ge 0,b\in {\bf Z}}\left\vert 1-
\bfe{kT+lU+bV}\right\vert^{4c(4kl-b^2)}
\end{equation}

(3) degenerate orbits
\begin{eqnarray}
  {\cal I}_{deg}&=&\frac{\pi}{3}c(0)U_2-\log Y^{c(0)}-\log
\prod_{l>0}\left\vert 1-\bfe{lU}\right\vert^{4c(0)}\nonumber \\[2mm]
&&\ \ +\left(\gamma_E-1-\log\frac{8\pi}{3\sqrt{3}}\right)c(0)
+4\pi\left(\frac{V_2^2}{U_2}+V_2+\frac{U_2}{6}\right)c(-1)\\[2mm]
&&\ \ -\log\prod_{l>0,b=\pm 1}\vert 1-\bfe{lU+bV}\vert^{4c(-b^2)}
-\log\prod_{b=-1}\vert 1-\bfe{bV} \vert^{4c(-b^2)}\nonumber
\end{eqnarray}
Here $U_2=Im U$, $V_2=Im V$ and $\gamma_E$ is the Euler-Mascheroni
constant.

Summing up the three contributions we obtain
\begin{equation}
  {\cal I}=-2\log \kappa Y^{10} \left\vert
\bfe{T+U+V}\prod_{(k,l,b)>0}(1-\bfe{kT+lU+bV})^{c(4kl-b^2)}\right\vert^2\,,
\end{equation}
where
\begin{equation}
  \kappa=\left(\frac{8\pi}{3\sqrt{3}}e^{1-\gamma_E}\right)^{10}
\end{equation}
and $(k,l,b)>0$ means that either $k>0,l\ge 0, b\in {\bf Z}$ or $k\ge
0,l> 0, b\in {\bf Z}$ or $k=l=0, b<0$.

Now a remarkable formula proved by Gritsenko and Nikulin \cite{rGNi}
shows that
\begin{equation}
  \label{prodformula}
  \chi_{10}(\Omega)=\bfe{T+U+V}\prod_{(k,l,b)>0}(1-\bfe{kT+lU+bV})^{c(4kl-b^2)}
\end{equation}
where
\begin{equation}
  \label{igusa}
  \chi_{10}(\Omega)=2^{-12}\prod_{\alpha:even}\{\theta_\alpha(\Omega,0)\}^2
\end{equation}
is the unique cusp form of weight 10 \cite{rIgusa}. Here
$\theta_\alpha(\Omega,0)$'s are theta constants of genus 2 and the
product is taken over the 10 even spin structures.  Actually
(\ref{prodformula}) is the ``square'' of the Gritsenko-Nikulin formula
\cite{rGNi}.

Thus finally we obtain
\begin{equation}
  \label{final}
{\cal I}=-2\log \kappa Y^{10}\vert \chi_{10}(\Omega)\vert^2.
\end{equation}
This precise form was recently proposed indirectly in \cite{rMS}\ by
demanding the existence of an infinite number of light Kaluza-Klein
particles in the decompactification limit \cite{rCFILQ}\ and that the
only logarithmic singularity is at $V=0$ where extra massless
particles appear\footnote{ The logarithmic singularity may easily be
  seen from the product formula (\ref{prodformula}).}.  The extra
massless particles are related to the vector multiplets needed for
symmetry restoration $U(1)\rightarrow SU(2)$ for $V=0$.

Notice that amusingly the argument of the logarithm in
eq.(\ref{final}) is the expression encountered in the two-loop
calculation of bosonic string \cite{twoloop} where the double zeroes
of $\chi_{10}$ at $V= 0$ were related to the existence of tachyon.

\section{Generalized Kac-Moody superalgebra and string-string duality}

In ref.\cite{rGNii}, Gritsenko and Nikulin introduced a generalized
Kac-Moody superalgebra without odd real simple roots for the Picard
lattice of an algebraic $K3$ surface $X$.  The Picard lattice is the lattice
of $H^{1,1}(X) \cap H^2(X,{\bf Z})$ and has signature $(1,n)$ where
$n\le 19$.  The Gritsenko-Nikulin formula \cite{rGNi}\ we used in this
work can be viewed \cite{rGNii} as the denominator formula of the
generalized Kac-Moody superalgebra associated with a certain Picard
lattice of signature $(1,2)$ generated by $\delta_1$, $\delta_2$ and
$\delta_3$ whose intersection matrix is
\begin{equation}
  (\delta_i\cdot\delta_j)=\left(
  \begin{array}{ccc}
     -2 &2 &2\\
      2 &-2&2\\
      2 &2 &-2
  \end{array}
\right)\,.
\end{equation}
$\{\delta_1,\delta_2,\delta_3\}$ is the set of even real simple roots
of the superalgebra.  More details can be found in the original
literature \cite{rGNi,rGNii}. See also related works \cite{rMartinec,
  rDolgachev}.  Obviously if we switch on more Wilson lines, we will
have to consider more general Picard lattices and the associated
superalgebra.

In ref.\cite{rHM} it is claimed that the vertex operators of vector
multiplets and hypermultiplets form a generalized Kac-Moody
superalgebra considering vector multiplets to be even and
hypermultiplets to be odd. The parity here is that of $b+1$ where $b$
is as in sect.2.  The shift by $1$ is due to the spectral flow between
the Ramond and Neveu-Schwarz sectors.  So it seems natural to consider
that the algebra is that of Gritsenko and Nikulin.

The input of our calculation of string threshold correction was the
$K3$ elliptic genus. The result was related to the Picard lattice of
an algebraic $K3$ surface and its associated generalized Kac-Moody
superalgebra thanks to the work of Gritsenko and Nikulin.  At first
sight, this is quite mysterious since mathematically remote objects
are correlated. From a physical point of view, the key to solve the
puzzle seems to reside in {\em string-string duality}.  As mentioned
in Introduction, when discussing $N=2$ heterotic-type IIA string
duality, the Picard lattices of algebraic $K3$ surfaces (the generic
fibers of $K3$ fibration Calabi-Yau threefolds) on the type IIA side
are related to the perturbatively visible gauge groups on the
heterotic side \cite{rAspinwall}.  (Aspects of symmetry enhancement of
type II string compactified on $K3$ have been discussed in
refs.\cite{rWittenii,rAspinwall,rBSV}.)  Thus it seems quite promising
to investigate string-string duality from the viewpoint of generalized
Kac-Moody superalgebras. In the work of Harvey and Moore \cite{rHM}
the application of their results to string-string duality has been
announced.

The content of this section is still speculative and apparently needs
further elaboration. I hope I can return to the issue in the near
future to substantiate the interpretation given here.

\section{Comments}

Even in the case of a single Wilson line that we considered in this
work the string threshold correction can in principle exhibit a more
complicated structure of moduli spaces \cite{rMS}. It would be
desirable to extend our world-sheet calculation in such a way that one
can see this directly.  It is also interesting to investigate the
heterotic compactification with more Wilson lines turned on and to
identify the Gritsenko-Nikulin superalgebras.

Another future direction would be to consider non standard-embedding
$N=2$ compactifications of heterotic string. As opposed to the cases
of Calabi-Yau threefolds, the $(0,2)$ elliptic genera of $K3$ do not
have much variety and are known \cite{rKM}.

What we have learnt in this work is restricted to the perturbative regime
on the heterotic side. It would be extremely interesting to
know how the picture will change non-perturbatively and how Seiberg-Witten
monopole (dyon) points appear.
On the type IIA side, the degenerate fibers of $K3$ fibrations are expected
to play a role \cite{rAspinwall}.

In any case, it is clear that there are many things we should learn
and clarify.

\end{document}